\documentclass[12pt]{iopart}
\usepackage[latin1]{inputenc}
\usepackage[english]{babel}
\usepackage[T1]{fontenc}
\usepackage{graphicx}
\usepackage{makeidx}
\usepackage{amssymb}
\usepackage{url}
\usepackage{xcolor}
\usepackage[bottom]{footmisc}

\newcommand{\MLcomment}[1]{}
\begin{document}
\title[Chirping signals in GW burst searches]
{Enhancing the significance of gravitational wave bursts through signal classification}
\author[S.Vinciguerra$^1$, M. Drago$^2$, G.A. Prodi$^3$, S. Klimenko$^1$, C.Lazzaro$^1$, V.Necula$^1$, G.A. Prodi$^1$, V.Tewari$^1$, M.Tringali$^1$, G. Vedovato$^1$]{S.Vinciguerra$^1$, M. Drago$^2$, G.A. Prodi$^3$, S. Klimenko$^4$, C.Lazzaro$^{5}$, V.Necula$^4$, F.Salemi$^2$, V.Tiwari$^6$, M.C.Tringali$^3$, G. Vedovato$^5$}
\address{$^1$ University of Birmingham, Edgbaston, Birmingham, B15 2TT,United Kingdom}
\address{$^2$ Albert-Einstein-Institut, Max-Planck-Institut f\"{u}r Gravitationsphysik, D-30167 Hannover, Germany}
\address{$^3$ Universit\'a di Trento, Dipartimento di Fisica, and INFN, TIFPA, I-38123 Povo, Trento, Italy}
\address{$^4$ University of Florida, Gainesville, Florida 32611, USA}
\address{$^5$ INFN, Sezione di Padova, I-35131 Padova, Italy}
\address{$^6$ Cardiff University, Cardiff CF24 3AA, United Kingdom}

\begin {abstract}
The quest to observe gravitational waves challenges our ability to discriminate signals from detector noise.
This issue is especially relevant for transient gravitational waves searches with a robust {\it eyes wide open} approach, the so called all-sky burst searches.
Here we show how signal classification methods inspired by broad astrophysical characteristics can be implemented in all-sky burst searches preserving their generality.
In our case study, we apply a multivariate analyses based on artificial neural networks to classify waves emitted in compact binary coalescences.
We enhance by orders of magnitude the significance of signals belonging to this broad astrophysical class against the noise background.
Alternatively, at a given level of mis-classification of noise events, we can detect about 1/4 more of the total signal population.
We also show that a more general strategy of signal classification can actually be performed, by testing the ability of artificial neural networks in discriminating different signal classes. 
The possible impact on future observations by the  LIGO-Virgo network of detectors is discussed by analysing recoloured noise from previous LIGO-Virgo data with coherent WaveBurst, one of the flagship pipelines dedicated to all-sky searches for transient gravitational waves.

\end {abstract}

\newpage\thispagestyle{empty}
\section{Introduction: signal classification for background rejection}
General searches for transient gravitational waves of generic waveform (GW {\it bursts}) have been accomplished exploiting the full sensitivity bandwidth of the Laser Interferometer Gravitational-Wave Observatory (LIGO) \cite{aasi2015advanced,Abbott:2007kv} and Virgo \cite{acernese2014advanced,Accadia:2012zzb} detectors. 
This type of \textit{all-sky} search has been performed by analyzing the network detector data with coherent methods \cite{burst-all-sky2010, burst-all-sky2012} looking for signals lasting from $ms$ to $s$ scale. These methods successfully identified the first detected gravitational wave, GW150914 \cite{detection, Burst}. 
In burst searches, the main factor which limits the statistical confidence of a gravitational wave candidate comes from non Gaussian noise outliers of single detectors, which may accidentally mimic a coherent response of the network. 
The implemented strategies to improve the capability of discriminating between signals and noise include both upstream and downstream methods. 
Upstream methods include data quality flags and vetoes at single detector level \cite{aasi2015characterization,abbott2016characterization} to clean the input of the network analysis. 
Downstream methods apply post processing procedures such as splitting the end results in a few separate frequency bands: to account for the most evident inhomogeneities of non Gaussian noise tails, any candidate belonging to a specific frequency band is thus ranked against the noise outliers characteristic of the same band.\\

Coherent WaveBurst (cWB) \cite{cWB} is the flagship pipeline aiming at all-sky burst searches on LIGO-Virgo data using minimal signal assumptions.
cWB has already been used for the analysis of data collected by the first generation of interferometers \cite{ MarcoTesi, ArXiv0050868v2} and during the first observation run of Advanced LIGO, O1.
In September 2015 cWB was the first pipeline to identify GW150914 \cite{Burst}.
cWB is based on a likelihood maximization of the coherent response of the network, which also allows the reconstruction of the most significant signal characteristics \cite{MarcoTesi,PRC}. 
To further reduce the false alarm probability at a reasonable cost in terms of false dismissals, additional procedures have been implemented. 
These procedures include simpler tests, such as the rejection of candidate signals in case of unusually high energy disbalance at different detectors, as well as more elaborate methods. 
Among them, cWB uses procedures for constraining the polarization and direction of detectable signals, since noise spectra and directional sensitivities affect the fraction of detectors which significantly contribute to the coherent response of the network.
All the strategies mentioned above preserve the degree of universality, typical of burst searches.\\

The performance of the pipeline can also take advantage of priors on the target signals; however, this is accomplished at the cost of a loss of generality of the search. 
Different versions of cWB pipeline have been tailored for, e.g., the search for coalescences of intermediate mass binary black holes \cite{IMBBHmethods} or of highly eccentric binary Black Holes \cite{eBBH}. \\

The scope of this work is to demonstrate how signal classification methods can complement all-sky burst searches to enhance the significance of selected signal classes without losing the generality of the search. 
The proposed signal classification method is based on machine learning techniques \cite{mashineLearning} to identify such signals against the non Gaussian noise outliers recorded in LIGO-Virgo observations. 
The signal vs. noise discrimination based on the use of machine learning techniques on single detector data has been discussed in several papers.
For instance \cite{IJMP24_11, Elena, zevin2016gravity} address the classification of transient noise outliers and \cite{CQG22_S1223} the whitening of the detector output. 
A signal recognition approach based on boosted decision trees has been tested for the case of a burst search triggered by astrophysical events in a network of detectors \cite{PRD88_062006}. 
Similarly, in \cite{Baker:2014eba}, the authors developed a multivariate classification with random forests in the context of matched filtering searches for high-mass black hole binaries.\\

The novelty of our work lies in its integration into all-sky burst searches and on the implementation of new strategies for signal classification, taking advantage of pattern recognition in the framework of the time-frequency (TF) representation of the candidate signals.
The latter is accomplished by averaging the response of more artificial neural networks (ANNs), as explained in Section \ref{SEC:method}.
As a case study, we focus on the recognition of signals consistent with coalescences of compact binaries (Section \ref{SEC:CaseStudy}).
Section \ref{SEC:Results} and Section \ref{SEC:Robustness} discuss respectively the effectiveness of the ANN analyses of the TF representations and its robustness against changes in the target signal distribution.
Section \ref{SEC:Multivariate analysis} explores the potentiality of a further multivariate step: we test a new ranking statistic built by combining the discrimination variables, adopted in a standard cWB analysis, with the output parameter describing the recognition of time-frequency patterns. 
By comparing the receiver operating characteristics, we show the achievable enhancement of GW burst searches through our signal classification strategies.
In Section \ref{SEC:FinalRemarks}, we give more general remarks about the impact of signal classification approaches on all-sky searches for GW bursts.

\section{Methodology}
\label{SEC:method}
With the purpose of enhancing the significance of GW bursts in all-sky searches, we 
complement the standard analysis pipeline cWB with a new discrimination variable based on ANNs, dedicated to signal classification according to their TF characteristics.
In its standard operation, cWB uses two main post-processing statistics to discriminate gravitational waves from noise artifacts (glitches): the network correlation coefficient ($cc$) and the effective correlated SNR ($\rho$) \cite{Burst}. The latter is used to rank candidate events and assign them a false alarm rate. The former measures the consistency of the candidate with a coherent response of the network to a GW.
cWB uses TF transformations \cite{JPCS363_012032} at different resolutions (or levels) and defines the TF characterisation for each candidate, through e.g. a Principal Component Analysis (PCA) \cite{cWBmanual}.
We use a PCA since it summarises the main TF characteristics of each candidate in the least number of TF pixels from different resolutions.

\begin{figure}[!hbt]
\centering
\includegraphics[width=16cm]{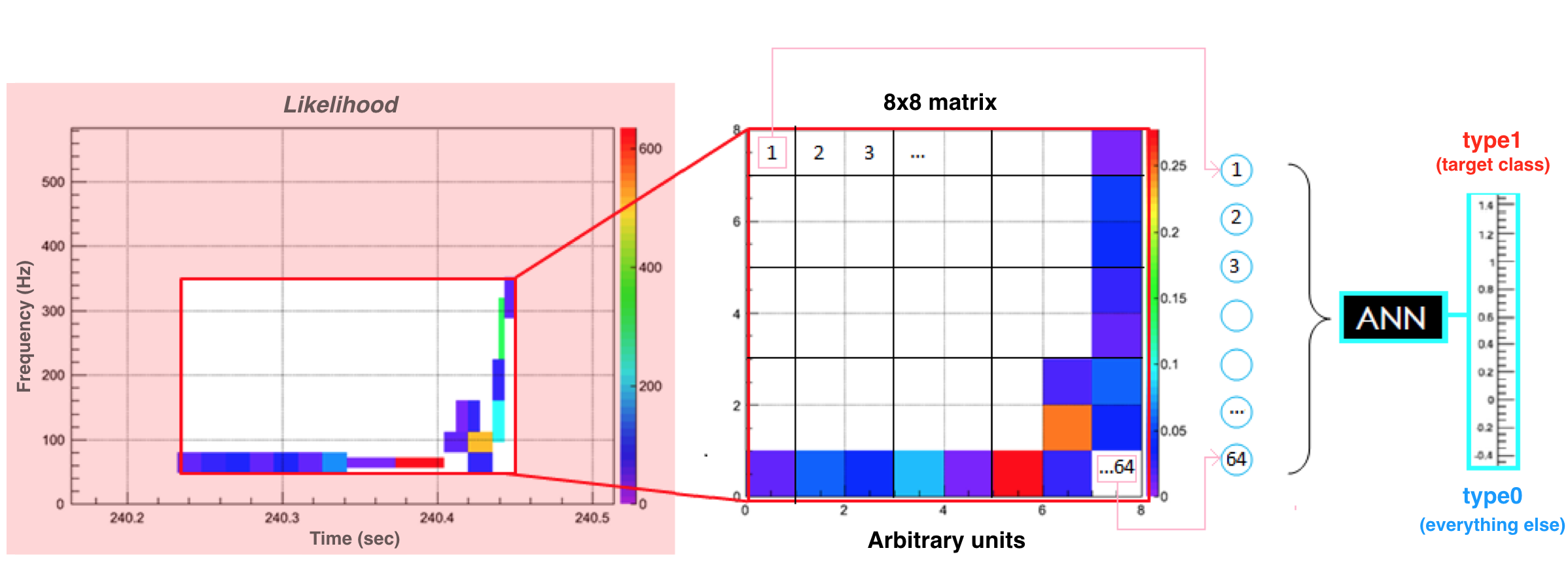}
\caption{Schematic representation of the proposed event classification. The procedure is based on the reconstructed time-frequency map of candidates. ANNs are trained to produce an output number close to $1$ for events are classified as belonging to the target distribution, and close to $0$ otherwise. Our procedure does not constrain the output value to be limited to [0,1] and overflows and underflows are possible.}
\label{Fig_tool}
\end{figure}

\begin{figure}[!hbt]
\centering
\includegraphics[width=12cm]{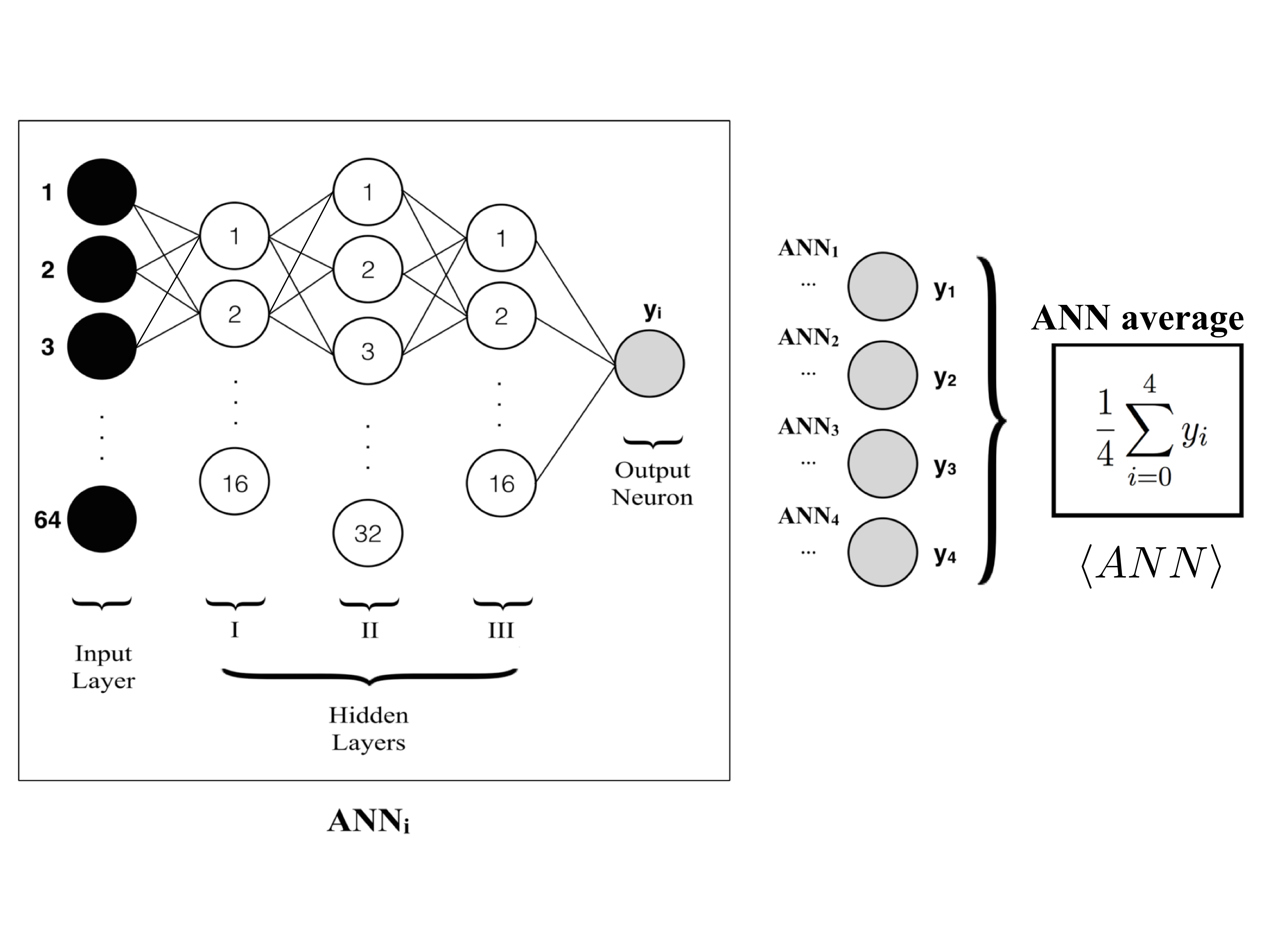}
\caption{
Schematic representation of the algorithm adopted to obtain the ANN average. In the square it is summarised the structure of each ANN: the black filled circles compose the input layer, while the empty ones represent elaborating units. Each of them sums the output values of the previous layer, by weighting them with the correspondent synapse's (lines) strength. 
The new classification parameter is obtained by averaging the result of 4 independent ANNs (ANN average).
}
\label{Fig_ANN}
\end{figure}

To perform the ANN classification, the resulting TF representations in the whitened data domain are converted by a post-processing algorithm into a frame made by $8\times 8$ pixels, each described by a scalar amplitude proportional to the fraction of total likelihood \footnote{\label{footN:1}In the algorithm dedicated to the definition of the frame, we also allow the rejection of late TF pixels with low likelihood. 
Starting from the pixels that appeared at the latest times, the procedure, which defines the $8\times 8$ frame, discards a fraction of the TF pixels, corresponding to at most 10\% of the total likelihood. 
This selection prevents the occurrence of problematic distortions of patterns in the $8\times 8$ frames.
Noisy pixels selected after the merger time would indeed shrink and move the characteristic trace left by chirping events on the $8\times 8$ frame. 
This operation is particularly useful for the classification in real noise. In fact, it favours a neater TF map by rejecting any weak structures appearing after the merger time, which will then dominate the last column of the frame (see {\it Fig \ref{Fig_tool}}).} (see {\it Fig \ref{Fig_tool}}).
Each frame carries no information about the absolute scale of the candidate strength, duration or frequency band and keeps only the uncalibrated shape information of the candidate's TF trace (for more details see \ref{App:TFtoFRAME}).\\
The 64 entries of the $8\times 8$ frame feed the ANNs dedicated to the classification of the TF pattern.
ANNs are informatic tools composed of calculation units (neurons - represented by circles in {\it Fig. \ref{Fig_ANN}}), connected together by synapses (represented by lines in {\it Fig. \ref{Fig_ANN}}), which acquire specific values (weights) accordingly to a supervised training procedure\footnote{A supervised training procedure consists in an optimization rule tailored to minimize the error between actual and desired output over the selected target and complementary sets of signals.} (see \cite{ANNurl} and references therein).
The ANNs adopted for this study are multilayer perceptrons defined within ROOT \cite{ANNmanual} (the object oriented framework developed at CERN).
Every ANN is composed of an input layer, fed by the $8\times 8$ frame, and of 3 hidden layers, requiring the definition of about $2\times 10^3$ synapses.
Each ANN is trained against a set of $\sim 10^4$ target signals (type1 class) and a set of $\sim 10^4$ complementary events (type0 class) composed of either noise glitches or signals belonging to alternative classes.
The classification rule consists in obtaining output values close to $1$/$0$ for elements of type1/type0 class respectively. 
For more details on the structure and training procedure used for the present study, see \ref{App1} and \cite{ANNmanual}.\\

\begin{figure}[!hbt]
\centering
\includegraphics[width=16cm]
{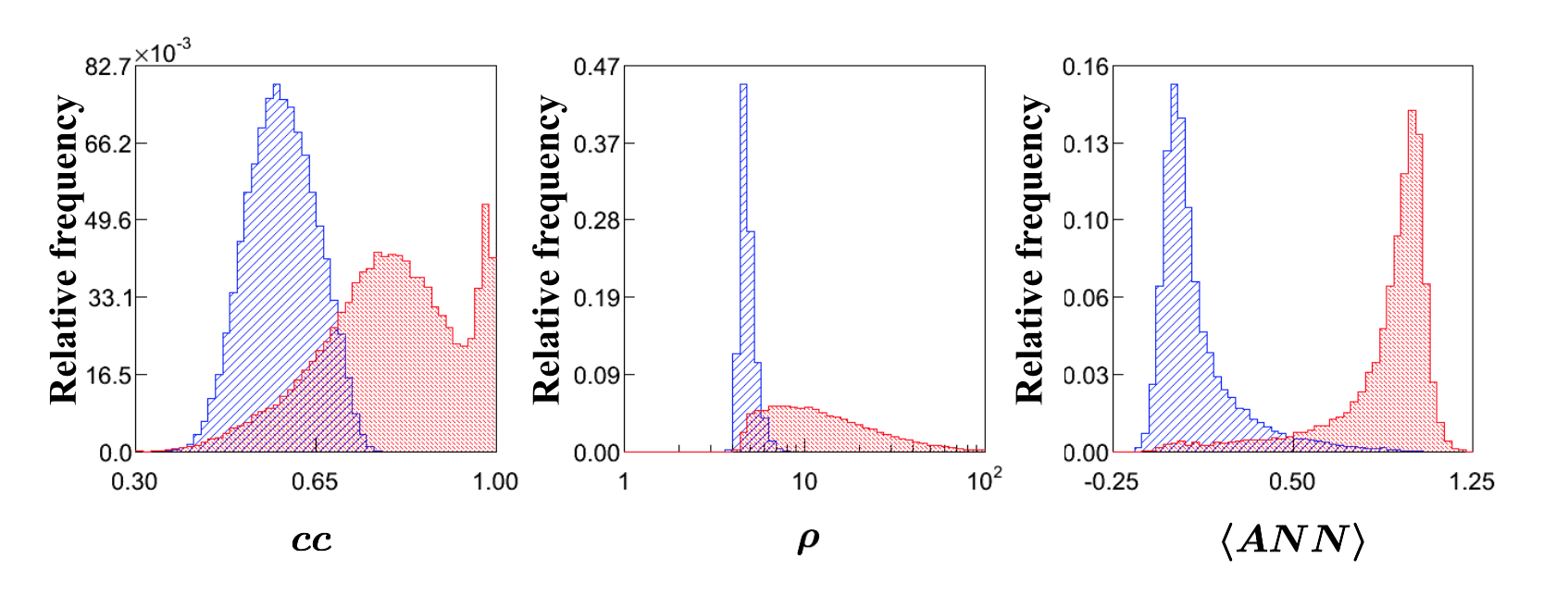}
\caption{Distributions of $5\times 10^4$ target CBC signals (red, see { \it Tab. \ref{tab3_1}}) and $5\times 10^4$ network glitches (blue) as seen in different analysis variables. \\ 
$y~\ axis$:  fraction of events in the $x$ bin.\\
$x~\ axis$, from left to right: network correlation coefficient, correlated SNR and average of 4 trained ANN outputs. The first two are the main test statistics used by cWB in GW burst searches. 
}
\label{Fis_istoTN}
\end{figure}

We mitigate the impact of the statistical fluctuations intrinsic to the ANN response by averaging the output of 4 (unless otherwise specified) independently trained ANNs and so introducing the {\it ANN average}, as shown in {\it Fig. \ref{Fig_ANN}}.
Every considered ANN is built with the same fixed training procedure, but each of them is trained over an independent set of events sampled from the same distributions.
The time required for the definition of a single ANN varies by several orders of magnitude according to the adopted training.
However, once defined, ANNs are able to quickly evaluate cWB triggers, providing an effective discriminating variable, as illustrated in {\it Fig. \ref{Fis_istoTN}}.
In this preliminary test, we considered a population of target signals (type1) made by compact binary coalescences (see { \it Tab. \ref{tab3_1}}) and a population of accidental coherent responses of the initial LIGO-Virgo network of detectors (network glitches, type0). 
The former is produced by software injections of signals in a few days of initial LIGO and Virgo data from S6D-VSR3 runs \cite{burst-all-sky2012,Aasi:2014mqd}, recolored according to the early phase spectral sensitivity of advanced detectors \cite{abbott2016prospects}. 
The set of network glitches is produced by running the standard cWB all-sky search on the same recolored data streams after applying a set of time-shifts among different detectors so to cancel any physical correlation present. The use of actual data is necessary to model the non Gaussian noise features that dominate the performances of all-sky burst searches. 
{\it Fig. \ref{Fis_istoTN}} shows that ANNs can be effective in discriminating these target signals from network glitches, to a much more efficient degree than the more general criteria used by cWB.
\section{Case study: classification of chirping signals}
\label{SEC:CaseStudy}
In this Section we focus on the classification of a specific signal class, emitted by compact binary coalescences. 
The results will be highlighted in Section \ref{SEC:Results} while their robustness will be discussed in Section \ref{SEC:Robustness}, proving that the enhancement provided by our classification still holds over a much wider signal parameter space than that used to train the ANNs. 
\subsection{Type1 or target signal class}
\label{SSEC:target}
\label{Simulation_Sec}
The results of the first observing run of advanced LIGO \cite{O1} and the measured rate for binary black hole (BBH) mergers of $9-240$ $Gpc^{-3}yr^{-1}$  \cite{Rates} confirm that compact binary coalescences are the most numerous sources for current ground based detectors. 
In this study, we focus on the classification of the inspiral phase of the binary evolution. 
During the inspiral, the gravitational emission is mainly determined by the chirp-mass $\mathcal{M}=\frac{\left(m_1 m_2\right)^{3/5}}{\left(m_1+m_2\right)^{1/5}}$, where $m_1$ and $m_2$ are the masses of the two companions.
The chirp-mass drives:
\begin{itemize}
\item the frequency evolution $\dot{f}\propto \mathcal{M}^{5/3}f^{11/3}$ of the gravitational waves, which sets the chirping behaviour commonly associated to these signals;
\item the gravitational wave strain amplitude, which in time domain  is:
$\mathcal{A}(t)\propto f^{2/3}\mathcal{M}^{5/3}$;
 \item the time spent in the most sensitive part of the spectral sensitivity of the detectors, which scales as $t_{det}\propto \mathcal{M}^{-5/3}$. This time is also strongly dependent on the noise spectral density (PSD) of detectors at lower frequencies.
\end{itemize}
Since the classification procedure depends on these properties, we investigated different distributions of chirp-masses.
In particular, we mainly tested two signal populations: 
\begin{itemize}
\item{a {\it low-mass} distribution (see \textit{Tab. \ref{tab3_0}}) composed of signals characterised by a clear chirping feature in the TF representation, which is dominating the detectable signal within the spectral sensitivity of the detectors.}
\begin{table}[hbt!]
\centering
\begin{tabular}{|c|c|}
\hline
\textbf{Mass distribution} & uniform in $M_{tot}$ and $m_1/m_2$  \\
\hline
\textbf{Total mass range [$M_{\bigodot}$]} & $M_{tot}\mathcal{2}[3,50]$ \\
\hline
\textbf{Mass ratio range}& $m_1/m_2\mathcal{2}[1,11]$ \\
\hline
\textbf{Distance range $d$ [Mpc]} & $\sim [70,225]$ \\
\hline
\textbf{number of Shells$^*$} \cite{Mazzolotesi} & $3$ \\
\hline
\textbf{Distribution in each shell} & uniform in volume \\ 
\hline
\end{tabular}
\caption{main parameters of the {\bf low-mass CBC signal distribution}.\\  
$^*$ The subdivision in shells is performed to decrease the computational load while ensuring more homogeneous statistical uncertainties on detection efficiency. The reference shell range is $[100-150]$~Mpc and 
by rescaling signal amplitudes we populate the two contiguous shells ($[\sim 70,100]$~Mpc and $[150,225]$~Mpc).}
\label{tab3_0}
\end{table}
\item{a {\it wide mass range} distribution (see \textit{Tab. \ref{tab3_1}}), composed of a CBC population including more massive systems and therefore shorter detectable signals, in which the inspiral phase plays a weaker contribution within the spectral sensitivity considered in this study (early phase of advanced detectors).}
\end{itemize}
\begin{table}[hbt!]
\centering
\begin{tabular}{|c|c|}
\hline
\textbf{Mass distribution} & uniform in $log(m_1)$, $log(m_2)$  \\
\hline
\textbf{Mass range [$M_{\bigodot}$]} & $m_{1,2}\mathcal{2}[1.5,96.0]$, $M_{tot}\mathcal{2}[3.0,136.0]$ \\
\hline
\textbf{Distance range $d$ [Mpc]} & $\sim [45,500]$\\
\hline
\textbf{number of Shells$^*$} \cite{Mazzolotesi} & $6$ \\
\hline
\textbf{Distribution in each shell} & uniform in volume\\ 
\hline
\end{tabular}
\caption{Main parameters of the {\bf wide-mass range signal distribution}. \\
$^*$ As described in \textit{Table \ref{tab3_0}}, but rescaling signal amplitudes to populate five contiguous shells from $\sim 45$~Mpc to $\sim 500$~Mpc. Detection efficiencies are listed in table \ref{tab3_2}.
}
\label{tab3_1}
\end{table}

\begin{figure}[!hbt]
\centering
\includegraphics[width=16cm]{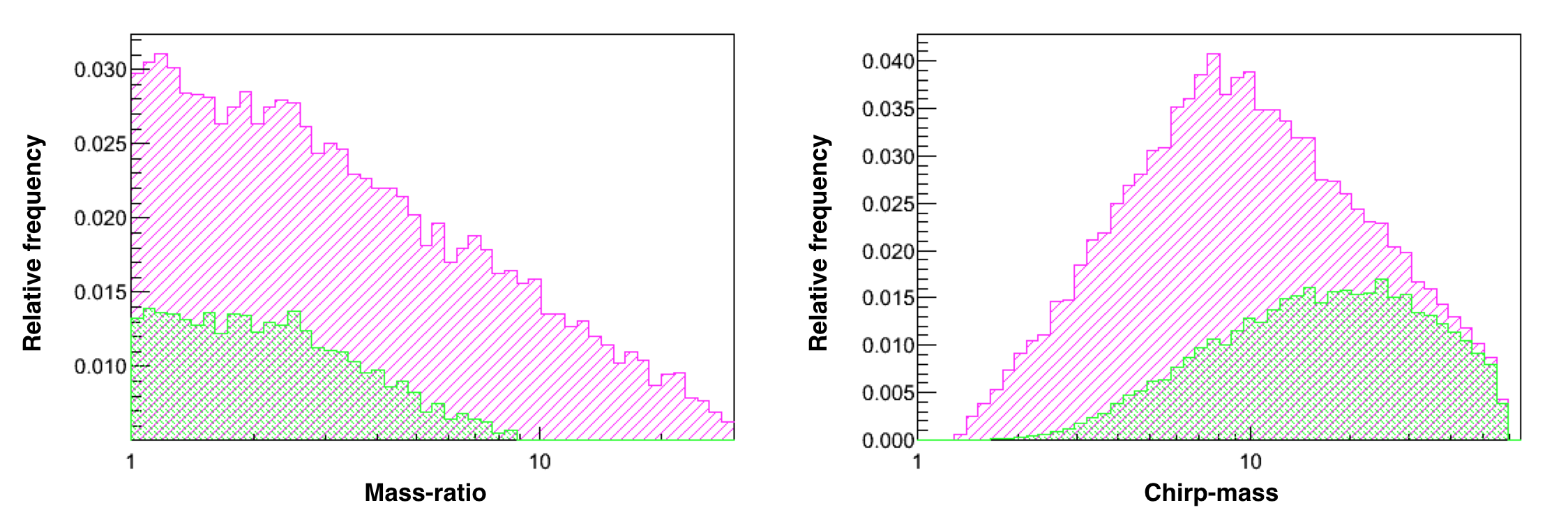}
\caption{Event distribution of injected signals (violet) and reconstructed by cWB (green)
for mass-ratio (left panel) and chirp-mass (right panel).
cWB pipeline reconstructed $\sim 6.5\times 10^4$ out of about $1.7\times 10^5$ signals, injected according to {\it Tab.\ref{tab3_1}}.
}
\label{Fig_SigDist}
\end{figure}

As CBC signals models, we adopt EOBNRv2 waveforms \cite{EOBNRv2Buon, EOBNRv297}.
They rely on the effective-one-body (EOB) formalism and describe all the phases (inspiral, merger and ring-down) of the coalescence.
The actual distributions of the detected signals used in this study are a convolution between the signal population (from {\it Tab. \ref{tab3_0}} or {\it Tab.\ref{tab3_1}}) and the cWB detection efficiency.
{\it Fig. \ref{Fig_SigDist}} illustrates the selection effect due to this convolution on the {\it wide-mass-range signal distribution} ({\it Tab.\ref{tab3_1}}). 
As expected the detection pipeline is more sensitive to louder signals, i.e. for more massive systems ({\it Fig. \ref{Fig_SigDist}}). Indeed the gravitational wave amplitude scales as $\mathcal{A}(t)\propto f^{2/3}\mathcal{M}^{5/3}$ and $\mathcal{M}(M,q)=Mq^{3/5}(q+1)^{-6/5}$, where $q = m_1/m_2$ is the mass ratio.
{\it Tab.\ref{tab3_2}} lists the overall detection efficiencies of cWB in the different shells. 
As expected, the pipeline efficiency decreases as the distance range increases.
\begin{table}[hbt!]
\centering
\begin{tabular}{|c|c|}
\hline
\textbf{Shell distance range [Mpc]} & \textbf{Efficiency \%} \\
\hline
[$\sim$45,$\sim$65] & $72.3 \pm 0.3$  \\
\hline
[$\sim$65,100] & $57.8 \pm 0.2$ \\
\hline
[100,150] & $42.5 \pm 0.1$ \\
\hline
[150,225] & $29.2 \pm 0.2$ \\
\hline
[225,$\sim$340] & $18.3 \pm 0.3$ \\
\hline
[$\sim$340,$\sim$505] & $10.2 \pm 0.1$  \\
\hline
\end{tabular}
\caption{cWB detection efficiency per shell \cite{Mazzolotesi}, for CBC signals belonging to the {\it wide-mass-range distribution} (see {\it Tab.\ref{tab3_1} and {\it Fig. \ref{Fig_SigDist}}})}
\label{tab3_2}
\end{table}
\subsection{Type0 or alternative signal class}
\label{SUBS:type0}
Our case study requires to test the classification procedure with respect to both network glitches and alternative GW signals.
The network glitch distribution has been described in Section \ref{SEC:method}. 
As alternative signal class we considered a mixture of GW signal waveforms widely used in simulations of GW burst searches, i.e. the BRST set described in \cite{abadie2012all}, which includes Gaussian pulses, sinusoidal signals with Gaussian amplitude envelope as well as White-Noise-Bursts waveforms. 
Such alternative signals lack of a proper astrophysical model and therefore their amplitude distribution at earth has been modeled by scaling their nominal amplitude value ($h_{rss}\equiv\sqrt{\int_{-\infty}^{\infty}h^2(t)\,dt} = 2.5\times 10^{-21}$) by a grid of logarithmically distributed scaling factors ($0.075$, $0.15$, $0.3$, $0.6$, $1.2$, $2.4$, $4.8$, $9.6$, $19.2$), see \cite{burst-all-sky2012,MarcoTesi}.
All these signals have been injected on the same recolored data set used for the target signals injections.
\section{Classification performance results}
\label{SEC:Results}
\label{Results_Sec}
The main three tests of classification performances are summarized in {\it Tab. \ref{tab:test1a}}.\\ 
\begin{table}[hbt!]
\centering
\begin{tabular}{|c|c|c|}
\hline 
& \textbf{type1 class} & \textbf{type0 class} \\
\hline
\textbf{Test 1} & chirp-like GWs from {\it low-mass distribution} (\it{Table} \ref{tab3_0}) &  network glitches  \\
\hline
\textbf{Test 2} & chirp-like GWs from {\it low-mass distribution} (\it{Table} \ref{tab3_0}) & alternative BRST GWs \\
\hline
\textbf{Test 3} & chirp-like GWs from {\it wide-mass distribution} (\it{Table} \ref{tab3_1}) & network glitches \\
\hline
\end{tabular}
\caption{Summary of type1 and type0 classes used for the main three tests of classification performances.}
\label{tab:test1a}
\end{table}

The results are described in terms of the fraction of type1 (type0) events which are correctly classified (miss-classified) as belonging to the target class, $F_{1\rightarrow 1} \ (F_{0\rightarrow1})$ defined by:
\begin{eqnarray}
F_{k\rightarrow 1} = \frac{1}{N_k}\sum_{i = 1}^{N_{k}}\delta_i
\end{eqnarray}
Here $N_k$ is the total number of tested events drawn from the $k$ class (where $k=1,0$ for type1,type0) and $\delta_i$ is defined for each event $i$ by:
\begin{eqnarray}
\delta_i = \left\{\begin{array}{lr}
        1, & \mathrm{if}~\rho_i\ge \rho_{th}~\land ~cc_i\ge cc_{th}~\land~ \left<ANN\right>_i\ge \left<ANN\right>_{th}\\
        0, & \mathrm{otherwise}
        \end{array}
        \right.
\end{eqnarray}
where the subscript ``${th}$'' refers to threshold value on the related variable. 
Events with $\delta_i=1$ ($\delta_i=0$) are classified as belonging to the type1 (type0) class. 
{\it Fig. \ref{Fig_S6Drec}-\ref{Fig_S6Dg_TN}}  summarise the results of each test and provide a comparison of the effects of the standard test statistics of cWB (correlated SNR and network correlation coefficient) with the ANN average.
Each figure is made by four plots, whose \textit{y-axis} reports $F_{1\rightarrow 1}$ and $F_{0\rightarrow 1}$ respectively in the top ones and bottom ones. 
The plots to the left show $F_{1\rightarrow 1}$ and $F_{0\rightarrow 1}$ as a function of the threshold on the correlated SNR, $\rho_{th}$. 
The red lines represent results obtained without applying a threshold on the ANN average (or $\left<ANN\right>_{th}=-\infty$), while green curves show the effects of selected threshold values, namely $\left<ANN\right>_{th}=\left\{0.0,0.5,1.0\right\}$.
The plots to the right show $F_{1\rightarrow 1}$ and $F_{0\rightarrow 1}$ as a function of the threshold on the ANN average. Here blue curves are computed by applying different thresholds on the correlated SNR $\rho_{th}=\left\{5,6,7\right\}$.
In all the plots a constant threshold on the network correlation 
coefficient is used, $cc_{th}=0.6$, which is a common choice in standard GW burst searches \cite{Burst}. 
This $cc$ threshold makes all the plotted fractions $F_{1\rightarrow 1}$ and $F_{0\rightarrow 1}$ lower than 1. 

\subsection{Test1: low-mass chirp-like GWs vs glitches}
\begin{figure}[!hbt]
\centering
\includegraphics[width=15cm]{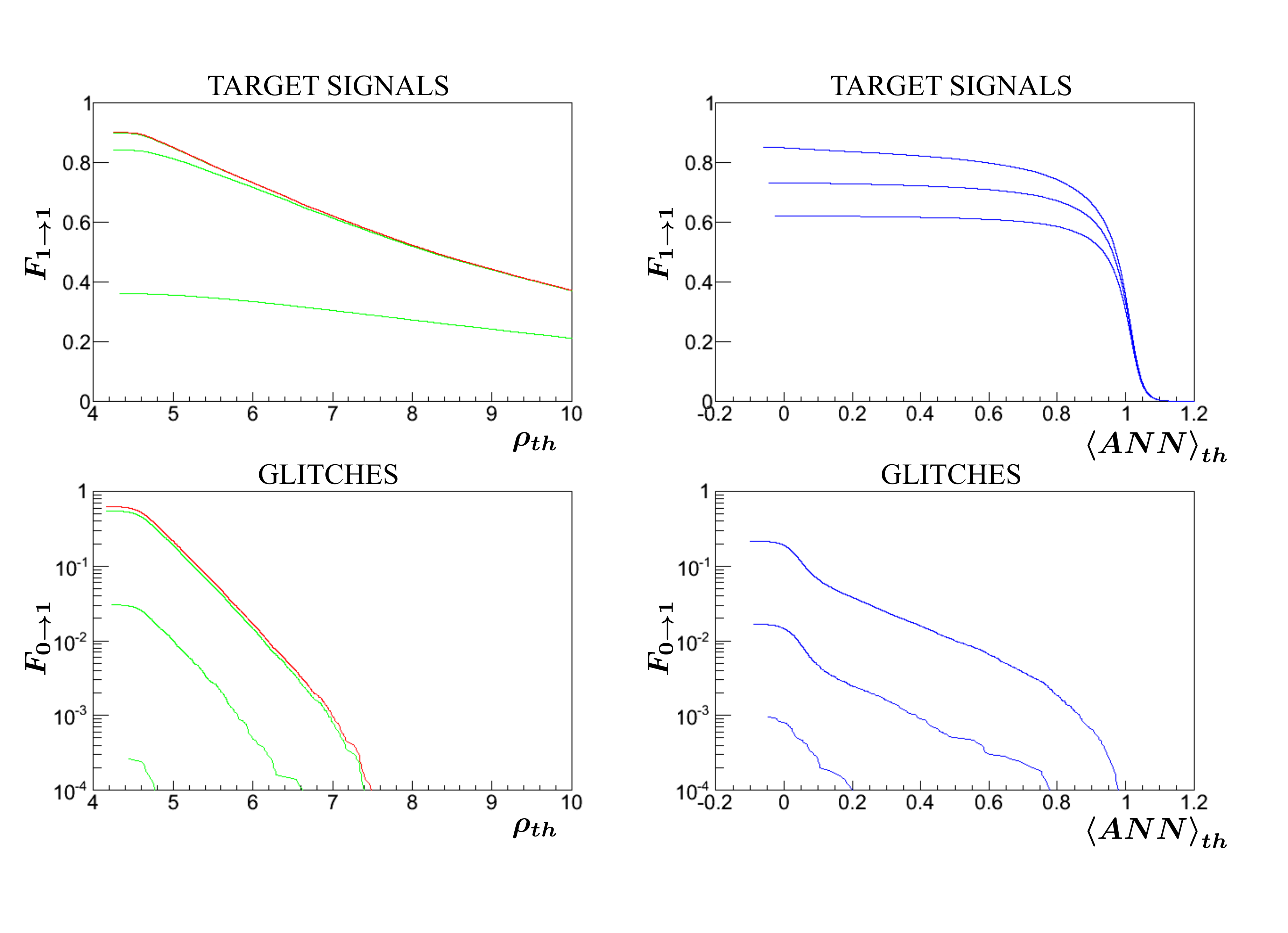}
\caption{
 Left plots: $F_{1\rightarrow 1}$ (top) and $F_{0\rightarrow 1}$ (bottom) as a function of $\rho_{th}$. 
Right plots: $F_{1\rightarrow 1}$ (top) and $F_{0\rightarrow 1}$ (bottom) as a function of $\left<ANN\right>_{th}$.
A constant threshold for $cc_{th}\ge 0.6$ is used. 
In the left plots, the red lines show the results obtained by a standard cWB analysis, while the green curves use also the signal classification with $\left<ANN\right>_{th}=\{0.0,0.5,1\}$, from top to bottom.
In the right plots, the blue lines are computed for $\rho_{th}=\{5,6,7\}$ from top to bottom.
The results refer to $5 \times 10^4$ events from the low-mass GW distribution and to the same number of network glitches.
}
\label{Fig_S6Drec}
\end{figure}

The first classification test aims to discriminate network glitches from chirp-like signals of the {\it low-mass} distribution (first line of table {\it Tab. \ref{tab:test1a}}).
The results on a population of $5 \times 10^4$ events per each type are reported in {\it Fig.~\ref{Fig_S6Drec}}: both correlated SNR and the ANN average are effective classifiers but their joint use shows advantages.
In fact, it is possible to enhance the statistical confidence with a much smaller cost in terms of detection efficiency.
For instance, considering the left plots, at any selected value of $\rho_{th}$, by adding $\left<ANN\right>_{th} = 0.5$ almost the same fraction of chirp-like signals are recovered, while the fraction of mis-classified glitches $F_{0\rightarrow 1}$ is reduced by about one order of magnitude.  
The right plots lead to similar considerations. $F_{1\rightarrow 1}$ depends very weakly on $\left<ANN\right>_{th}$ as long as $\left<ANN\right>_{th} \leq 0.8$, while the mis-classified fraction of events $F_{0\rightarrow 1}$ drops substantially in the same $\left<ANN\right>_{th}$ range.

\subsection{Test2: low-mass chirp-like GWs vs alternative GWs}
\begin{figure}[!hbt]
\centering
\includegraphics[width=15cm]{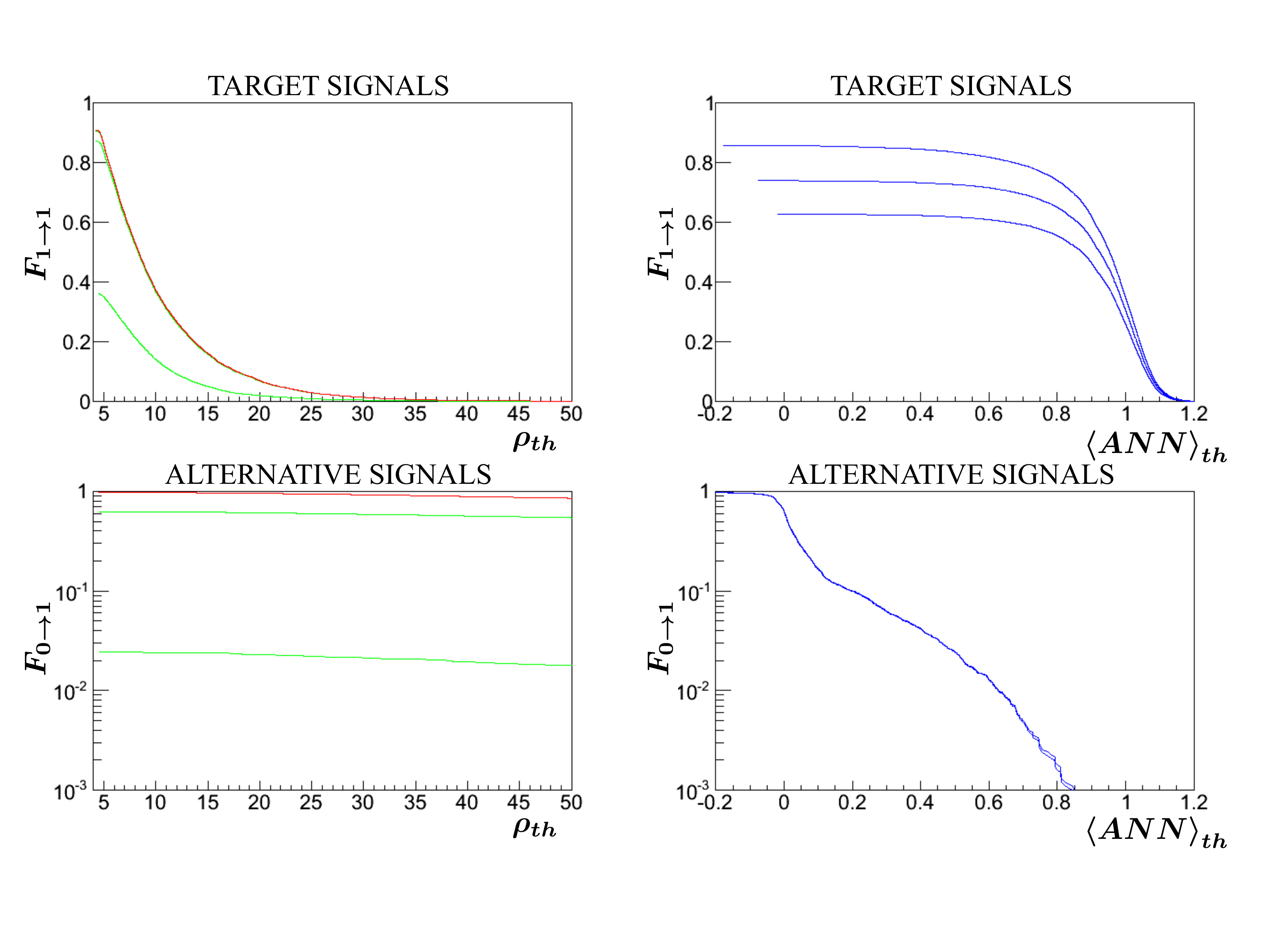}
\caption{
The Figure is structured as {\it Fig. \ref{Fig_S6Drec}}. The right plots clearly illustrate that the ANN average provides an efficient separation of the two GW populations. 
Results refer to $10^4$ type1 signals from the {\it low-mass} distribution and an equal number of type0 events from the alternative signal class, formed of BRST simulation. 
All the signals have been injected in recoloured detectors data.
For this test, the ANN average is computed from the outputs of just 2 ANNs.}
\label{Fig_BRSTTest}
\end{figure}

The signal classes considered for this test are the {low-mass} distribution and the BRST simulation set introduced as alternative signal class in \ref{SUBS:type0}.
The results of the classification are reported in {\it Fig. \ref{Fig_BRSTTest}}, which shows the performances in terms of $F_{1\rightarrow 1}$ and $F_{0\rightarrow 1}$ with the same structure of {\it Fig. \ref{Fig_S6Drec}}.
In particular, the right plots illustrate that the ANN average provides an efficient separation of the two GW populations, while the correlated SNR is instead agnostic with respect to the GW waveform class. 
In this test we selected a population of alternative GWs which are louder than the target GWs, to test an opposite condition with respect to what described in the previous subsection. The results are consistent with the ANN average being agnostic with respect to the loudness of the events, as it was designed to be.
The chosen loudness of the alternative GW class carries no specific physical meaning, and the difference between $F_{1 \rightarrow 1}$ and $F_{0 \rightarrow 1}$ in the left plots cannot be used to classify type1 and type0 events in an actual GW search.

\subsection{Test3: wide-mass chirp-like GWs vs glitches}
\begin{figure}[!hbt]
\centering
\includegraphics[width=15cm]{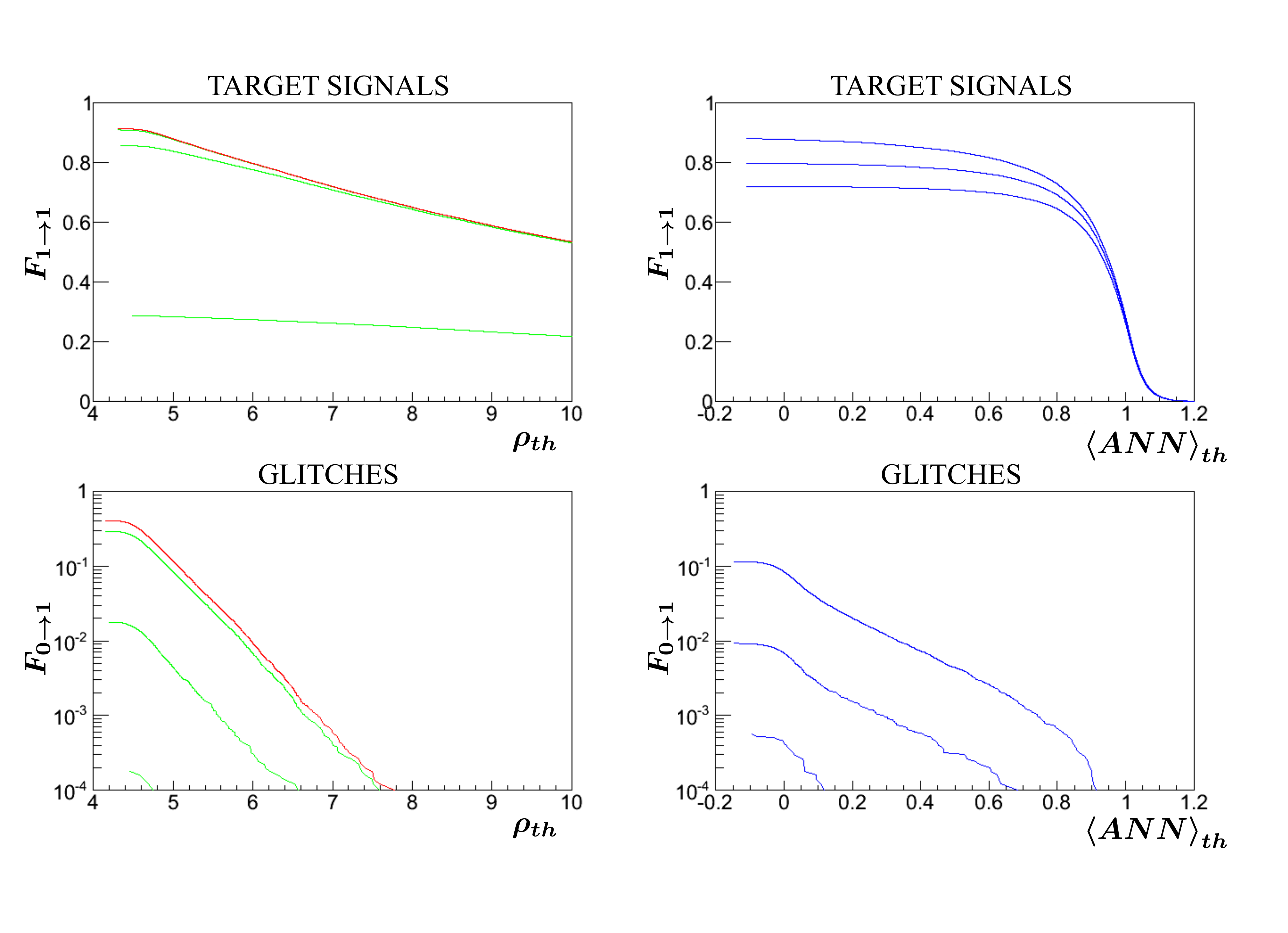}
\caption{
The image is structured as {\it Fig. \ref{Fig_S6Drec}}. 
The test is applied on $5\times 10^4$ type1 signals from the {\it wide-mass} range distribution and $5\times 10^4$ type0 events from the set of recolored network glitches.
}
\label{Fig_S6Dg_TN}
\end{figure}

The task of the last classification test is separating network glitches from chirp-like signals, drawn from the {\it wide-mass} range simulation ({\it Table \ref{tab3_1}}). 
We report the results in {\it Fig. \ref{Fig_S6Dg_TN}}.
The plots of {\it Fig. \ref{Fig_S6Dg_TN}} and {\it Fig. \ref{Fig_S6Drec}} exhibit very similar trends. 
The main difference is that $F_{1\rightarrow 1}$ values are generally slightly higher, since the {\it wide-mass} event distribution includes also louder GWs.
Instead, even if the chirping character is weaker for the wide-mass signals distribution, the results show that the ANNs can be trained to give comparable performances with respect to the  low-mass case. 
\section{Robustness}
\label{SEC:Robustness}
\label{Rob_Sec}
\begin{figure}[!hbt]
\centering
\includegraphics[width=15cm]{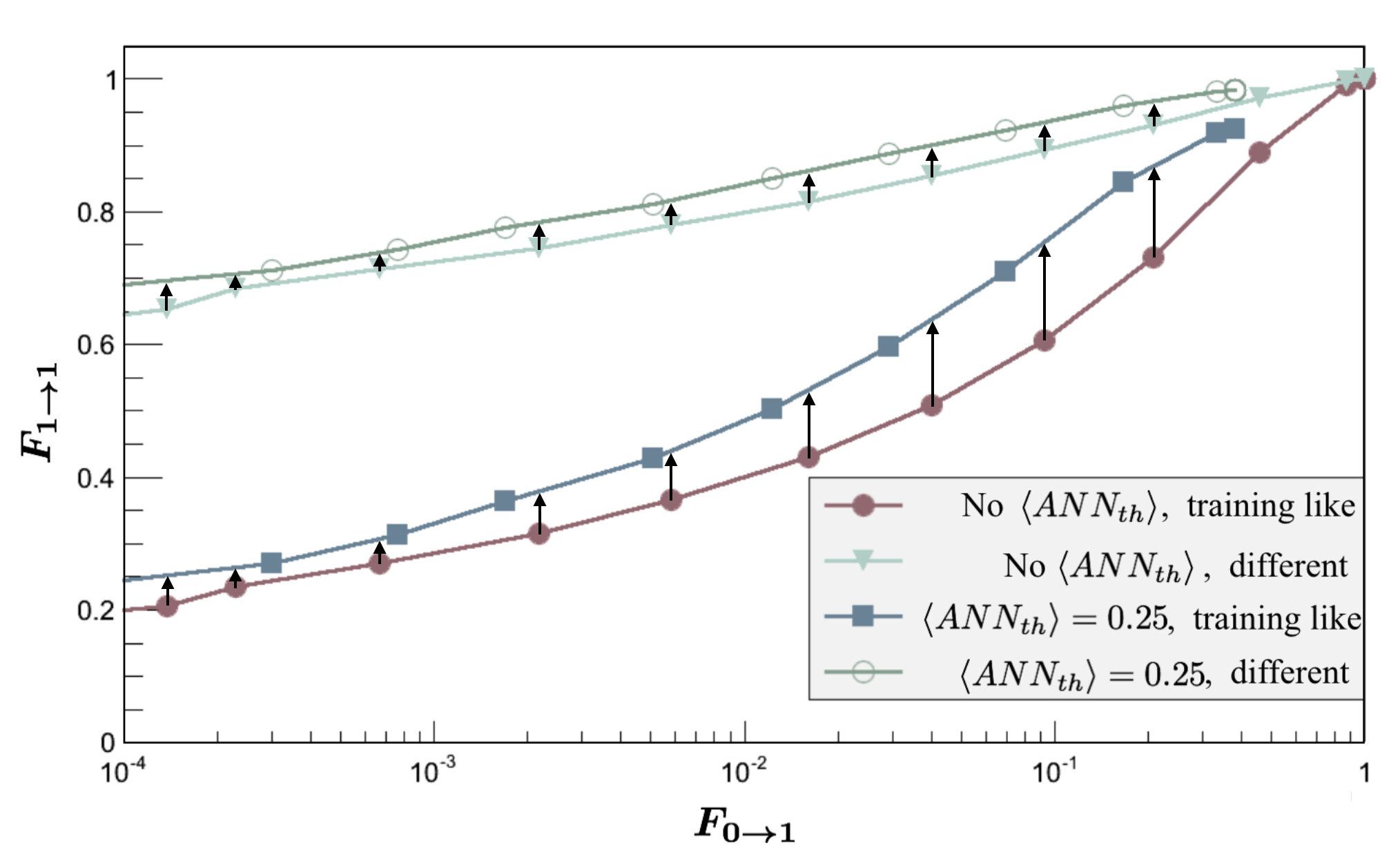}
\caption{
$F_{1\rightarrow 1}$ vs $F_{0\rightarrow 1}$ measured with (blue squares and green void cycles) and without (red filled circles and teal triangles) the application of a threshold on the ANN average of $0.25$. Symbols correspond to the same grid of $\rho_{th}$ values.
The red and blue curves ({\it matched}) are the results performed by testing the same signal population used to train the ANNs. 
The green and teal lines ({\it mismatched}) trace the $\overline{ROC}$ for a different test set built selecting CBC signals from {\it the wide-mass} range distribution. 
In both the cases, $5\times 10^4$ CBC signals form the type1 class and the same number of network glitches form the type0 one. 
}
\label{Fig_ROC_comp_summary}
\end{figure}

When the learning is supervised \cite{ANNurl}, as in our case, the ANN structures are defined thorough procedures mainly driven by the selected type0 and type1 samples.
Since the astrophysical distribution of chirp-like GWs is unknown, we need to investigate the robustness of our approach against biases in the training distributions.
To this purpose, we consider Receiver Operating Curves ($\overline{ROC}$) computed as $F_{1\rightarrow 1}$ vs $F_{0\rightarrow 1}$ by varying the value of $\rho_{th}$, while keeping constant $cc_{th}=0.6$, at selected $\left<ANN\right>_{th}$ values. 
{\it Fig. \ref{Fig_ROC_comp_summary}} shows $\overline{ROC}$ curves for $\left<ANN\right>_{th} = -\infty$ and $\left<ANN\right>_{th} = 0.25$.
The robustness is tested by comparing the improvement related to the application of the $\left<ANN\right>_{th} = 0.25$ on the $\overline{ROC}$
in two different cases:
\begin{itemize}
\item classification results are performed on chirp-like signals belonging to the same type1 distribution used for the training;
\item classification results are performed on chirp-like signals belonging to a distribution different from that of the type1 used for the training.
\end{itemize}
\begin{description}
\item[Training set]
We build a simulation of GW emissions from {\it three fixed equal-mass binaries} $(10-10)M_{\bigodot}$, $(25-25)M_{\bigodot}$ and $(50-50)M_{\bigodot}$ (see {\it Table \ref{3sys_dist}}).
From each of these three sources, we collect an equal number of reconstructed events to compose the training set for the target (type1) signals.
The type0 training class is again defined with a sample of network glitches recoloured to mimic the spectral sensitivity in the early phase of the advanced detectors.\\
\item[Testing sets]
The first one is composed of samples of the same type1 and type0 populations used to define the training set just described, while the other one is formed by events drawn from the type1 and type0 classes of line 3 in table {\it \ref{tab:test1a}}.
For both cases, the tested type0 events are independent samples of the recoloured network glitches. 
\end{description}
As expected, the $\overline{ROC}$ curves for a testing set different from the training set are worse, see {\it Fig.} \ref{Fig_ROC_comp_summary}.
However, the improvement on the $\overline{ROC}$ achieved by implementing the threshold on the ANN average are significant and of comparable value in both testing sets.
In this sense, {\it Fig. \ref{Fig_ROC_comp_summary}} demonstrates the robustness of our approach against biases in the population-model.
Since the mismatched type1 testing set has a much wider range of chirp-masses than the matched one, we present the achieved results of {\it Fig. \ref{Fig_ROC_comp_summary}} as representative of the effect 
produced by increasing the assumed volume in the parameter space of the source population.\\
\section{Multivariate analysis}
\label{SEC:Multivariate analysis}
\label{MVA_Sec}
The last versions of the cWB pipeline include an estimation of the binary chirp-mass $\mathcal{M}_{est}$, computed as a best fit of the signal TF trace resulted from the PCA.
To compute the fit, the algorithm considers all the pixels obtained with the TF decomposition at different levels, discarding the ones flagged as noise artifacts, according to the procedure defined in \cite{Burst, CHIRP-MASS}.
Applying such post-processing analysis significantly improves the signal to noise discrimination achieved by cWB for CBC signals \cite{CHIRP-MASS}.\\
\begin{figure}[!hbt]
\centering
\includegraphics[width = 15cm]{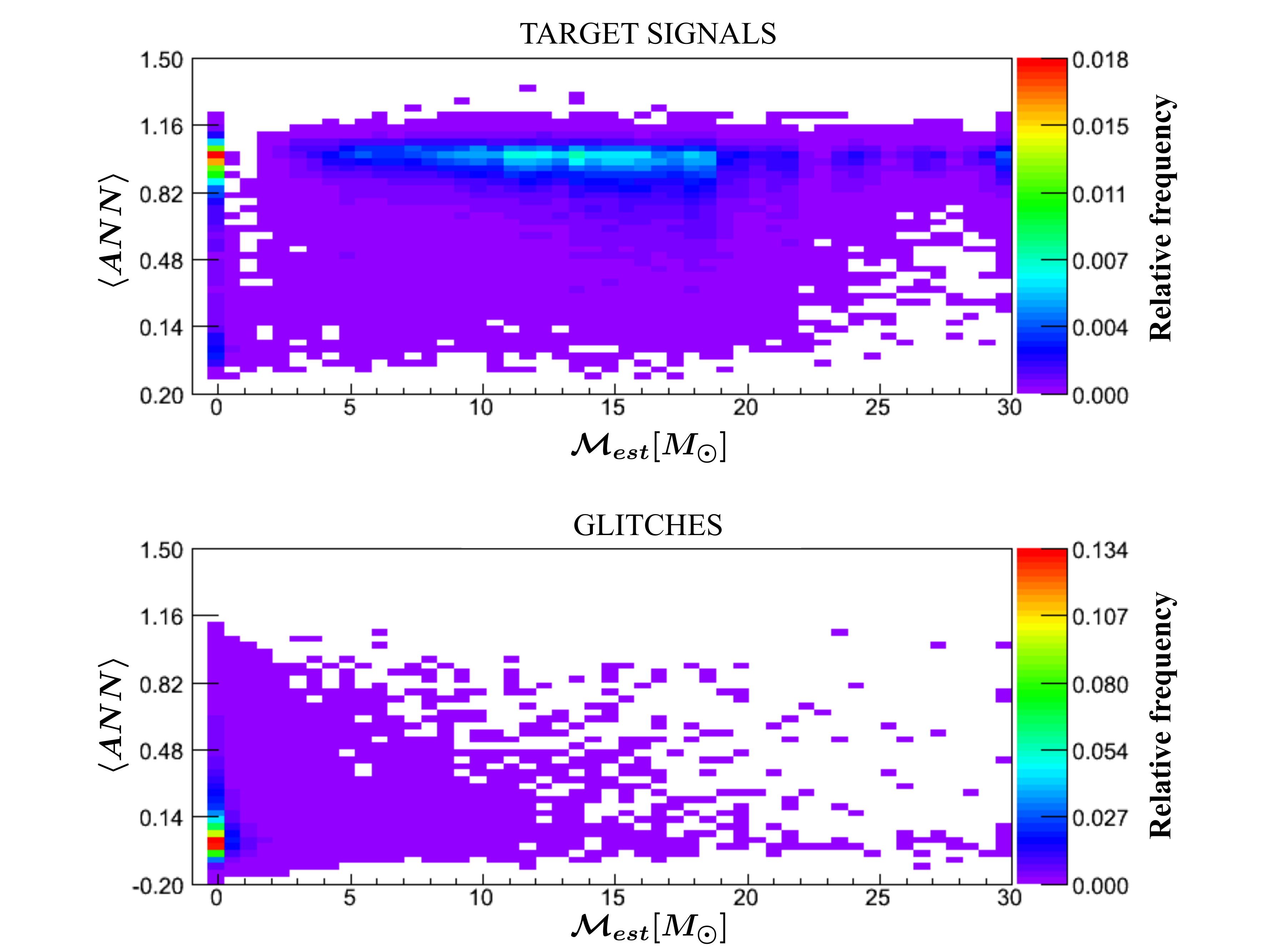}
\caption{
Distributions on plane defined by ANN average and the estimated chirp-mass of candidates belonging to {\it the wide-mass} signal class ({\it Top panel}) and S6D recoloured glitches ({\it Bottom panel}). The colour scale represents the fraction of recovered events, over a total of $5\times 10^4$ per class. 
Negative values estimated chirp-mass are automatically set to zero. 
}
\label{Fig_ComparisonMcANN}
\end{figure}
The joint distribution of the estimated chirp-mass and the ANN average (see {\it Fig. \ref{Fig_ComparisonMcANN}}) demonstrates that the two variables are not fully correlated. 
This gives the opportunity to further improve the results by a joint use of these variables.
In addition we can consider other signal parameters estimated by cWB to implement a multivariate analysis (MVA).\\

\begin{figure}[!hbt]
\centering
\includegraphics[width=12cm]{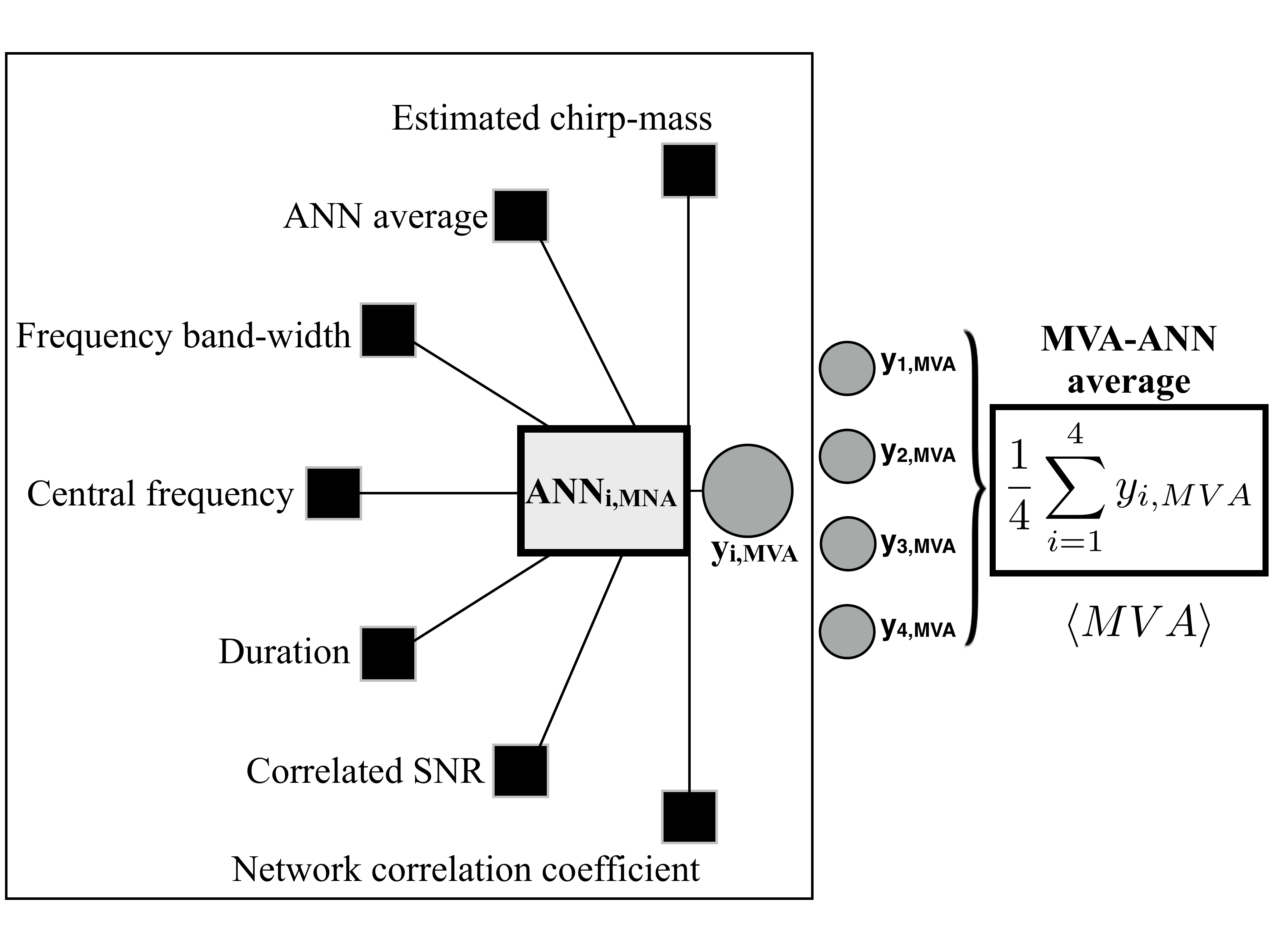}
\caption{Schematic representation of the algorithm adopted to define the MVA-ANN average. In the left square, we list all the input quantities elaborated by all the four $\mathrm{ANN}_{\mathrm{i,MVA}}$ to obtain $\left<MVA\right>$.
}
\label{fig:MVA}
\end{figure}

For the morphological discrimination we are interested in, we implement an  additional classification stage, using four MVA-ANNs (see {\it Fig. \ref{fig:MVA}}).
They are characterised by approximately 300 synapses and independently trained on $\sim10^3$ events per class. 
The input quantities listed in {\it Fig. \ref{fig:MVA}} are then used to define a MVA-ANN average ($\left<MVA\right>$).
We evaluated the effectiveness of the multivariate approach by comparing the $\overline{ROC}$ obtained from this MVA ranking statistic (MVA-ANN average) with the one driven by different values of correlated SNR, accordingly to the standard cWB analysis.
\begin{figure}[!hbt]
\centering
\includegraphics[width=15cm]{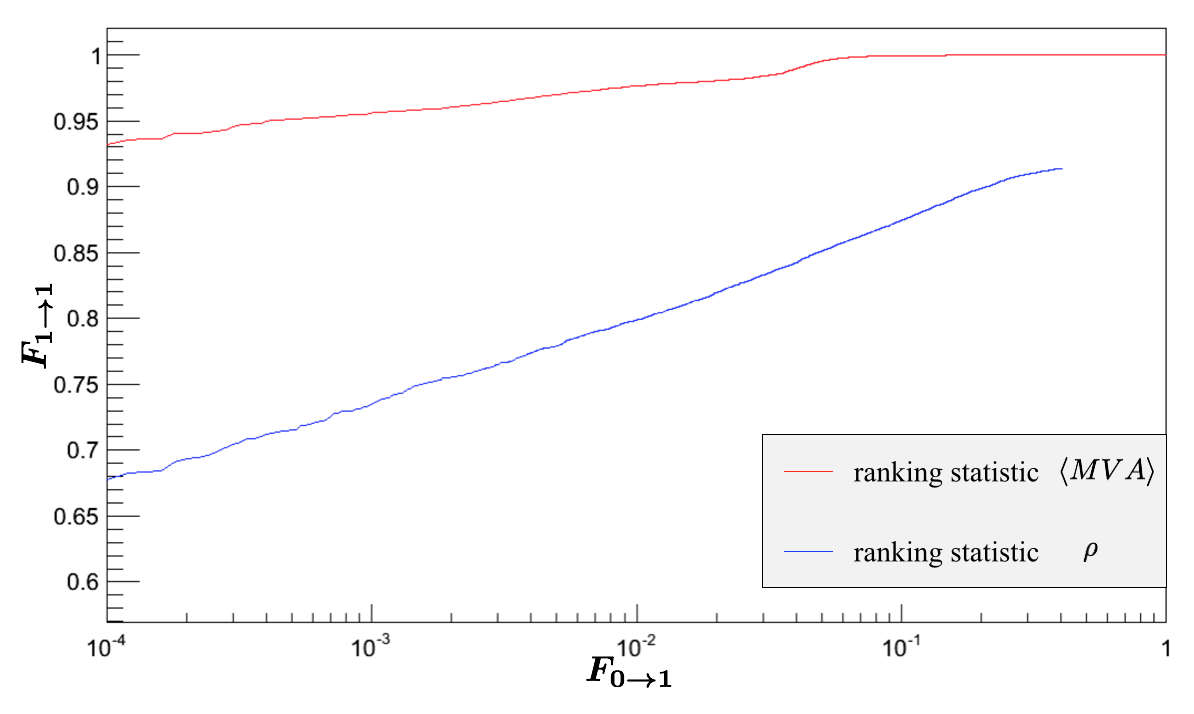}
\caption{$\overline{ROC}$ using the MVA-ANN average (red) and the correlated SNR (blue) as a ranking statistic.
In the last case, $F_{1\rightarrow1}$ and $F_{0\rightarrow 1}$ are obtained considering a constant threshold on the network correlation coefficient $cc_{th}>0.6$ and ignoring the ANN average. 
The testing set, as well as the training one, has been defined by $5\times 10^4$ samples of S6D recoloured glitches (type0) and by $5\times 10^4$ signals drawn from the {\it wide-mass} range distribution (type1).
}
\label{GenANN}
\end{figure}
{\it Fig. \ref{GenANN}} clearly demonstrates that cWB's performances in discriminating signals from glitches can be considerably improved by adopting the MVA-ANN average as ranking statistic.
In fact, at fixed $F_{1\rightarrow1}$ values $F_{0\rightarrow1}$ is lowered down by three orders of magnitude, when switching the ranking statistic from the correlated SNR to the MVA-ANN average.\\

We finally performed checks to point out the sensitivity of our MVA-ANN average
to the different inputs.
These tests show that the estimated chirp-mass and the ANN average are by far the ones that impact more the results. 
Much smaller contributions comes from the central frequency, the correlated SNR and the network correlation coefficient, while duration and frequency bandwidth are the parameters which appear to be the least effective for our multivariate analysis.
\section{Final Remarks}
\label{SEC:FinalRemarks}
In the previous sections we presented the application of the signal classification to the selected case study, i.e. CBC-like transient signals with an inspiral stage chirping up in frequency. 
The resulting enhancement of the significance of the detected signals in this class is improved by orders of magnitude, at least in the confidence range investigated here (see {\it Fig. \ref{GenANN}}). 
Alternatively, the gain can be interpreted as a significant increase of the detected fraction of sources at a given confidence, e.g. we recover $\sim 25\%$ more signals at a mis-classification of 0.01\% of noise events. The current approach could be further improved by considering more candidate parameters ($cc$, $\rho$, $\mathcal{M}_{est}$, etc ), as proposed in Section \ref{MVA_Sec} by defining a new ranking statistic $\left<MVA\right>$ and by generalising the classification to other signal classes.\\

This search can be easily integrated within the framework of an all-sky search, where the search is general and open to every kind of GW-like signals with no particular assumptions on the morphology. 
In this situation, we can split the overall set of interesting triggers, $W$, in two classes, the say {\it class $A\subset W$} of CBC-like signals and the complementary {\it class} $(W - A)$ of triggers belonging to $W$ but not to $A$. 
Following this approach for many waveform classes, all-sky searches can be managed as more separate searches on the same observation time, similarly to the more traditional case of all-sky searches performed on separate frequency bands. 
To account for the increased number of trials, a relative weight on these searches has to be chosen to portion out the overall false alarm probability of the all-sky search.\\

Signal classes of astrophysical interest can be, for example, ring-down-like signals as emitted in quasi normal modes of NSs \cite{NSsignals} - say {\it class $B$} signals, or, more generically, signals with a reconstructed duration longer than a  number of typical cycles, allowing for a diversity of waveform amplitude envelopes and phase evolutions in time - say {\it class $C$} signals. 
In addition, glitch classification methods can also be implemented for vetoing purposes, i.e. to reject the more frequent noise transient families at the detectors - say {\it class $Z$}. 
The most straightforward implementation of the all-sky search would then be a hierarchical signal classification scheme in subsequent steps, such as e.g. selecting $A\subset (W-Z)$, then $B\subset (W-Z-A)$, then $C \subset(W-Z-A-B)$ and finally analysing the rest $(W-Z-A-B-C)$.
Such a hierarchical implementation would ensure that the considered classes are disjoint, both for  signal and for glitches. 
Of course the more background rejection is accomplished in the first classification steps, the higher will be the resulting background left to the last ones. \\

In the framework of all-sky searches, the optimization strategy is not defined, both because we are lacking reliable source population models for most source classes and because we want to be leave room for unexpected detections. 
The prioritization of signal classes and the portioning of the overall false alarm probability among the classes is subjective and has to be agreed upon, as a balance between boosting detection probability of better known sources and preserving suitable detection chances for signals in the widest accessible duration-frequency range. 
The former requires to take into account the detectable source number within the visible volume of the search; therefore, it would prioritize e.g. the frequency band of best spectral sensitivity and/or some waveform or polarization class. 
The latter instead calls for the consideration of the entire spectral range of the detectors, including disadvantaged spectral sensitivity bands, and for unmodelled waveforms and polarization states. \\

This issue has to be addressed anyway, regardless of the implementation of signal classification methods. 
In past all-sky searches, the portioning of the false alarm probability has been driven by uniform priors, i.e. by accounting a-posteriori for the trial factor coming from multiple sub-searches on different bandwidths or, in other words, by ranking signal candidates according to a quantity closely related to their inverse false alarm rate, as measured within the related sub-search. 
A portioning close to uniform makes sense also for the hierarchical search depicted here. \\

\section*{Acknowledgment}
We would like to thank Ilya Mandel for the useful suggestions concerning this work.\\
The research leading to these results has received funding from the 
People Programme (Marie Curie Actions) of the European Union's Seventh 
Framework Programme FP7/2007-2013/ (PEOPLE-2013-ITN) under REA grant 
agreement n \textdegree [606176]. It reflects only the author's view and that the 
Union is not liable for any use that may be made of the information 
contained therein.

\appendix
\section{Conversion of TF representations into $8\times 8$ frames}
\label{App:TFtoFRAME}
ANNs are here trained to recognise the common patterns in the TF representation of chirping signals. 
ANNs input layer are feed with the values of an $8\times 8$ frame constructed starting from the Time-Frequency Principal Component Analysis run by cWB 2G. 
The TF representation is defined by applying the Wilson-Daubechies-Meyer (WDM) transformation to the data \cite{necula2012gravitational} at different time ($\Delta T$) and frequency ($\Delta F$) resolutions (related to each other by $\Delta T \times \Delta F = 1/2$). 
Then the algorithm applies the PCA to the most energetic pixels of each map ({\it core pixels}) to represent a particular event. 
From this multiple map representation, we analyse the possibility of discarding pixels, according to the rule mentioned in the first footnote of Section \ref{footN:1}. 
With this information we are able to focus (\textquotedblleft zoom\textquotedblright) on the TF region really involved by the event. This selected region is divided into fundamental units, i.e. the minimum time resolution of the selected core pixels and half of the minimum frequency resolution (according to the application of WDM transformations). 
We finally group all the resulted units so to obtain a $8\times8$ frame. 
In order to adjust the \textquotedblleft zoomed\textquotedblright{}  region to the $8\times8$ frame,  each of the $8\times8$ squares can contain more fundamental units, or a fraction of them. 
The corresponding $64$ values are therefore obtained by summing or spreading the likelihood of the all fundamental units used to define each frame square.
These values are then normalised and used to feed the ANNs. 
More details can be found in \cite{ANNmanual}.

\section{Analysis and parameters}
\label{App1}
\textit{Table \ref{tab_cwbparameters}} reports the cWB-parameters adopted for the transient searches described in the paper.\\
\begin{table}[hbt!]
\centering
\begin{tabular}{|c|c|c|c|}
\hline
PARAMETER & VALUE & PARAMETER & VALUE\\
\hline
$\rho$ threshold & $5$ & cc threshold & $\sim 0.5$\\
\hline
search type & {\it i} & detector network & V1H1L1\\
\hline
range of f & from $64Hz$ to $2048Hz$ & data set & S6D\\
\hline
range of $\delta t$ & from $\sim3.90ms$ to $250ms$ &  range of $\delta f_{max}$ & from $2Hz$ to $128Hz$ \\
\hline
$\Delta t_{cluster}$ & $3$ {\it s} & $\Delta f_{cluster}$ & $130${\it Hz}\\
\hline
\end{tabular}
\caption{Main cWB parameters used to analyze the recolored data \cite{CQG25_114029}, \cite{JPCS363_012032}.}
\label{tab_cwbparameters}
\end{table}

The training procedure, applied to construct the ANNs used for classifying TF patterns, is defined by the parameters reported in \textit{Table \ref{tab_ANNpar}} (for more details on the parameters and their choice see \cite{RootMLP} and \cite{ANNmanual}).
\begin{table}[hbt!]
\centering
\begin{tabular}{|c|c|}
\hline
\textbf{Training set} & 16384 BKG-events, 16384 SIG-events\\
\hline
\textbf{Epoch number} & 650 \\
\hline
\textbf{Normalization} & to the total of each matrix representation\\
 & to the maximum for each matrix element on the training set\\
\hline
\textbf{Architecture} & IN: 64; H: 16/32/16; OUT: 1\\
\hline
\textbf{Learning method} &  Conjugate Gradients with F.R. updating formula\\
\hline
\end{tabular}
\caption{The training set includes the examples used for the preliminary tests. The architecture describes the input (IN), the hidden (H) and the output (OUT) layers through their numbers of neurons.}
\label{tab_ANNpar}
\end{table}
\pagebreak
\section{Other signal distribution}
To test robustness of the proposed analysis against the uncertainty over the chirp-like signal distribution, we introduced another class  of chirping signals, whose distribution is defined by the parameters reported in {\it Tab. \ref{3sys_dist}.
\begin{table}[hbt!]
\centering
\begin{tabular}{|c|c|}
\hline
\textbf{Mass range [$M_{\bigodot}$]} & $m_{i}\mathcal{2}\{10,25,50\}$ $i\mathcal{2}\{1,2\}, m_1=m_2$ \\
\hline
\textbf{Mass distribution} & uniform in the 3 $m_{i}$ values  \\
\hline
\textbf{Distance range $d$ [Gpc]} & $\sim[10^{-4},R_{f}], R_f\mathcal{2}\{0.7,1.5,2.6\}$\\
\hline
\textbf{Distance distribution} & uniform in $d^3$ \\
\hline
\end{tabular}
\caption{Main parameters and correspondent values adopted to construct the distribution of chirp-like events  used in {\it Sec. \ref{SEC:Robustness}}.
$R_f$'s values are calculated taking into account the relation $SNR \propto \frac{\mathcal{M}^{5/6}}{d}$.
}
\label{3sys_dist}
\end{table}
\newpage
\section*{Bibliography}
\bibliographystyle{unsrt}
\bibliography{biblio.bib}

\end{document}